\documentclass[twocolumn,prl,showpacs]{revtex4}

\usepackage{graphicx}
\usepackage{dcolumn}%
\usepackage{bm}%
\usepackage{amssymb}

\def\be{\begin{equation}}
\def\ee{\end{equation}}

\def\lan{\langle}
\def\ran{\rangle}

\begin{document}

\title{Dynamic heterogeneities in attractive colloids}
\author{A. Fierro$^{a,d}$, E. Del Gado$^{b}$, A. de Candia$^{a,c,d}$,
and A. Coniglio$^{a,e}$}
\affiliation{${}^a$ Dipartimento di Scienze Fisiche, Universit\`a di
Napoli ``Federico II'',\\ Complesso Universitario di Monte
Sant'Angelo, via Cintia 80126 Napoli, Italy}
\affiliation{${}^b$ ETH Z\"urich, Department of Materials, Polymer Physics,
CH-8093 Z\"urich, Switzerland}
\affiliation{${}^c$
CNISM Universit\`a di Napoli ``Federico II''}
\affiliation{${}^d$ INFN Udr di Napoli}
\affiliation{${}^e$ INFM CNR Coherentia}
\date{\today}
\begin{abstract}

We study the formation of a colloidal gel by means of Molecular
Dynamics simulations of a model for colloidal suspensions. A slowing
down with gel-like features is observed at low temperatures and low
volume fractions, due to the formation of persistent structures. 
We show that at low volume fraction the dynamic susceptibility, which
describes dynamic heterogeneities, exhibits a large plateau, dominated by
clusters of long living bonds. At higher volume fraction, where the effect
of the crowding of the particles starts to be present, it crosses over
towards a regime characterized by a peak.
We introduce a suitable mean cluster size of clusters of monomers
connected by ``persistent" bonds which well describes the dynamic 
susceptibility.

\end{abstract}


\maketitle
Recent advances in colloidal science allow to obtain colloidal
particles or nanoparticles with
specific functional properties, of electronic, chemical, biological
or mechanical nature.
Hence the packing and aggregation of colloidal particles are
important for a wide variety of applications, including biological
arrays, sensors, paints, ceramics, and photonic crystals.

In particular, by adding polymers to a colloidal solution, it is
possible to induce an effective short range attraction between colloidal
particles, known as depletion effect.
Attractive colloids exhibit a rich phenomenology in the
temperature - volume fraction plane \cite{edinburgh, weitzna,
weitzclu, dinsweitz, nature, weitz, attr_exp}: At high
temperature the attraction can be neglected and, by increasing
the volume fraction, if crystallization is avoided, a hard sphere
glass transition occurs.
By decreasing the temperature, the effect of the short range
attraction induces an attractive glass line which is strongly
temperature dependent, as predicted by Mode Coupling Theory
\cite{attr}, and by mean field theory \cite{caiazzo}.
This line at low volume fraction is identified as a gelation line.
In fact at low temperature and low volume fraction, attractive
colloids are known to exhibit a structural arrest with properties
similar to the sol-gel transition \cite{weitzclu, sciortino1,
kroy, physicaa, sciortino2}.
Although intensively studied both
experimentally and numerically, the theoretical understanding of
colloidal gelation, compared with chemical (irreversible) gelation
and glass transition, is still far from being reached.

A promising approach to the comprehension of the complex dynamics of
such systems is based on the study of the dynamical heterogeneities.
This concept has been successfully introduced in
glasses \cite{cicerone, kob, franz, bennemann, berthier, biroli, glotzer},
to take into account the correlated motion of particle clusters,
which decorrelates after a characteristic time $\tau$, of the order of the
relaxation time. These dynamic heterogeneities are described
quantitatively via the so-called dynamic susceptibility
\cite{franz}.
In a Lennard-Jones binary mixture, a typical model
for glass transition \cite{kob-and},
the dynamic susceptibility grows as a function
of the time, reaches a maximum and then decreases to a constant
\cite{glotzer}, consistently with the transient nature of the
dynamic heterogeneities.
Dynamic heterogeneities have been also observed in attractive
colloidal systems in both experiments \cite{weeks,cipelletti} and
numerical simulations \cite{attr_sim,reichman,charb}. In particular
in \cite{charb} a systematic study of the dynamic susceptibility was
done along the attractive glassy line. Typically the dynamic
susceptibility displays a well pronounced peak, however in
the attraction-dominated limit, the dependence on both time and wave
vector markedly differs from that in standard repulsion-dominated
systems (hard-sphere limit).

A rather different behavior was instead found recently in a model
for irreversible gels, made of monomers with permanent
bonds \cite{tiziana_prl}. It was shown in fact that
the dynamic susceptibility, defined as the fluctuations of the self
Intermediate Scattering Function (ISF) \cite{biroli}, in the limit
of low wave vector, $k\rightarrow 0$, tends for long times to a
plateau, whose value coincides with the mean cluster size. As a
consequence, as the system approaches the gel transition (i.e. the
percolation threshold), the value of the plateau diverges.
Such finding shows that, in irreversible gelation, the
heterogeneities coincide with clusters of monomers linked by
permanent bonds, and differently from glasses,
have a static origin and do not exhibit any decay.

In this paper we study the dynamic susceptibility in a DLVO type of
model for colloidal gelation \cite{israel,bartlett}, at very low
volume fraction and temperature, where the system exhibits more
markedly gelation properties. Using MD simulations we find an
interesting and completely unusual behavior.  The dynamic
susceptibility, for low wave vectors, 
increases with time until it reaches a large plateau value of
the order of the mean cluster size, as found in the model for irreversible
gels, and for time larger than the bond lifetime, it decays to 1.
Only at higher volume
fractions, where the bond lifetime is comparable to the
density-density structural relaxation time, the dynamic
susceptibility exhibits a crossover towards the glassy behaviour with
a well pronounced peak. 
Interestingly, a proper definition of a time dependent mean
cluster size, where a cluster is made of ``mobile'' particles connected by
bonds present at both time $0$ and
$t$, well reproduces the observed behavior of the dynamic susceptibility
for a fixed low temperature and low volume fractions. At higher volume
fraction, however, the dynamic susceptibility starts instead to display a
peak and shows a discrepancy with the time dependent mean cluster size.  This
indicates a crossover towards a new regime where, besides the clusters, also
the crowding of the  particles starts to play a role in the slowing down of
the dynamics.
Although we have considered a DLVO type of model, we expect that the
main results of our paper should be valid also for other model
systems exhibiting colloidal gelation
\cite{emanuela-walter,sciortino-valence}.

{\em The model - } 
The DLVO model, considered here, has been previously studied using
Molecular Dynamics (MD) \cite{physicaa, sciortino2,
noi_tubi}. In agreement with experimental findings
\cite{bartlett}, this model displays a structural arrest very close
to the percolation threshold at low temperature, where clusters are
made of particles connected by long living bonds \cite{nota_bond}.

In Refs. \cite{noi_tubi}, we found
that, at low temperatures, increasing the volume
fraction, the
system undergoes a transition from a disordered cluster phase to an
ordered hexagonal lattice of tubular structures and,
at higher volume fraction, to an
ordered lamellar phase. If this ordered state is
avoided, the system enters a ``supercooled'' metastable liquid phase
until structural arrest (gel) occurs \cite{physicaa, marco} very close to the
percolation threshold.
For this reason we introduce here  a small degree of
polydispersity, which actually hinders the formation of the ordered
phases. In this way we are able to fully investigate and
characterize the gel formation at different temperatures and
volume fractions.
The interaction potential between two particles $i$ and $j$ is:
\begin{equation}
V_{ij}(r) = \epsilon \left[A \left(\frac{\sigma_{ij}}{r}\right)^{36} -
B\left(\frac{\sigma_{ij}}{r}\right)^6 +
C\frac{e^{-r/\xi}}{r/\xi}\right]
\label{eq_dlvo}
\end{equation}
where $A=3.56$, $B=7.67$, $C=75.08$, $\xi=0.49$ \cite{nota1},  and
$\sigma_{ij}=(\sigma_i + \sigma_j)/2$.
In the following we consider the radii $\sigma_i$
randomly distributed in the interval
$\sigma-\delta/2  <  \sigma_i  < \sigma+\delta/2$ with $\delta=0.05 \sigma$.
The potential is truncated and shifted to zero at a distance of
3.5$\sigma$.
The temperature $T$ is in units of $\epsilon/k_B$, where
$k_B$ is the Boltzmann constant, wavevectors are in units of $\sigma^{-1}$,
and times in units of $\sqrt{m\sigma^2 /\epsilon}$, where $m$ is the mass of the particles.
The volume of the simulation box is kept constant, $V=5000\pi\sigma^3/3$, and
different volume fractions are obtained by varying the number of particles $N$,
so that $\phi= \pi \sigma^3 N/ 6 L^3$.
We have performed Newtonian MD at constant NVT using the velocity Verlet
algorithm and the
Nos\'e-Hoover thermostat with time step $\Delta t=0.01$.

At low temperatures
and low volume fractions, we find the same properties as in the
monodisperse case \cite{physicaa}: A cluster phase followed by a percolation
transition in the same universality class as random percolation.

\begin{figure}
\begin{center}
\includegraphics[width=7cm]{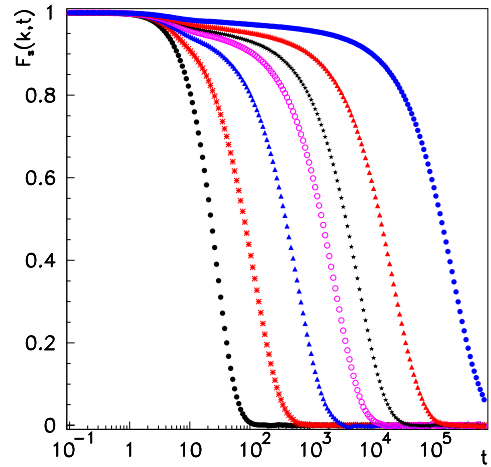}
\end{center}
\caption{(color online).
The self ISF, $F_s(k,t)$, for $T=0.15$, $k_{min}=~2\pi/L\simeq 0.36$, and
$\phi=~0.01$, $0.05$, $0.08$, $0.10$, $0.11$, $0.12$, $0.13$
(from left to right).}
\label{fig1}
\end{figure}

{\em Self Intermediate Scattering Function-}
In order to study the dynamic behaviour of the model, we measure the
self ISF,  $F_s(k,t)=\frac{1}{N}\lan \Phi_s(k,t)\ran$, where
$\Phi_s(k,t)= \sum_{i=1}^N e^{i\vec{k}\cdot(\vec{r}_i(t)- \vec{r}_i(0))}$,
$\langle \dots \rangle$ is
the thermal average, and $\vec{r}_i(t)$ is the position of the $i$-th particle
at the time $t$.
In Fig.~\ref{fig1} $F_s(k,t)$ is plotted for $T=0.15$ and
wave vector $k_{min}=2\pi/L\simeq 0.36$.
Increasing the volume fraction, the relaxation functions show a
two-step decay,
with a plateau value decreasing with the wave
vector; The long time tail is fitted by stretched exponentials.

\begin{figure}
\begin{center}
\includegraphics[width=7cm]{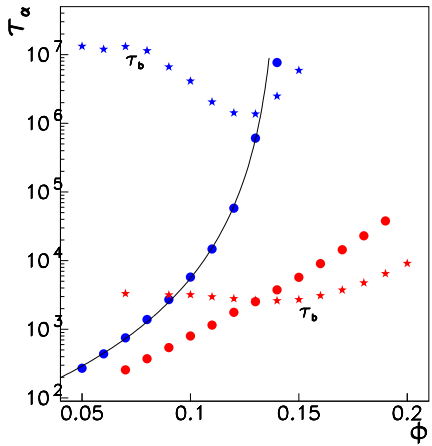}
\end{center}
\caption{(color online).
The structural relaxation time, $\tau_\alpha(k_{min})$ (circles),
compared with the bond relaxation time, $\tau_b$ (stars), for $T=0.15$ and
$0.25$ (from bottom to top). The continuous line is a power law fit
$(0.14-\phi)^{-3.8}$.}
\label{fig2}
\end{figure}

From the self ISF we calculate the structural relaxation time,
$\tau_\alpha(k)$, defined as $F_s(k,\tau_\alpha(k))\simeq 0.1$. In
Fig.~\ref{fig2} $\tau_\alpha(k_{min})$ is plotted as a function of the
volume fraction for two values of temperature, $T=0.15$ and $0.25$.
In the same figure for the sake of comparison we also plot the bond lifetime,
$\tau_b$ \cite{nota}, defined through the bond correlation function as
$B(\tau_b)\simeq 0.1$,
where $B(t)=\frac{\sum_{ij}\left[ \langle
n_{ij}(t)n_{ij}(0)\rangle-\langle n_{ij} \rangle^2\right]}
{\sum_{ij}\left[\langle n_{ij}\rangle-\langle
n_{ij}\rangle^2\right]}$, $n_{ij}(t)=1$ if particles $i$ and
$j$ are bonded at time $t$, $n_{ij}(t)=0$ otherwise
\cite{nota_bond}.
The data show that, at the lower temperature, $T=0.15$, and at low
volume fractions, where the bond lifetime is larger than the
structural relaxation time, $\tau_\alpha(k_{min})$ can be fitted by a
power law, $(\phi_c-\phi)^{-a}$. In this regime the system
dynamically behaves as if made of permanent clusters as in
irreversible gels. At the higher temperature, $T=0.25$, instead
the bond lifetime is of the same order of magnitude than the
structural relaxation time, and we observe a smooth increase of
$\tau_\alpha(k_{min})$ not associated to a power law critical
behavior.

{\em Dynamic heterogeneities -} According to the above
interpretation of the dynamics, we should expect, for time
windows comparable or larger than $\tau_b$, a deviation from the
dynamics of irreversible gels even for $T=0.15$. This crossover in
the dynamics can be better stressed by looking at the dynamic
susceptibility, defined as
$\chi_4(k,t)=\frac{\rule{0pt}{10pt}1}{N}\left [\lan
|\Phi_s(k,t)|^2\ran-|\lan \Phi_s(k,t)\ran|^2\right ]$.

\begin{figure}
\begin{center}
\includegraphics[width=7cm]{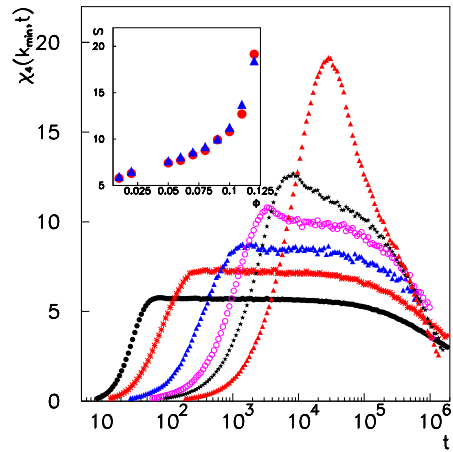}
\end{center}
\caption{(color online). {\bf Main frame}: The fluctuations of the self
ISF, $\chi_4(k_{min},t)$, for $T=0.15$ and
$\phi=~0.01$, $0.05$, $0.08$, $0.10$, $0.11$, $0.12$ (from left to right).
{\bf Inset}:
The mean cluster size, $S$ (triangles), as a function of the volume
fraction, $\phi$, compared with the maximum of the dynamic
susceptibility, $\chi_4(k_{min},t^*)$ (circles).}
\label{fig3}
\vspace{-0.5cm}
\end{figure}

In the main frame of
Fig.~\ref{fig3}, $\chi_4(k_{min},t)$, is plotted as function
of $t$, for
$T=0.15$ and different  $\phi$. For low volume fractions, after a
time roughly of the order $\tau_\alpha$, $\chi_4(k_{min},t)$ reaches
a plateau, close to the value of the mean cluster size,
$S=\sum_s s^2 n_s/\sum_sn_s$ (see also
the inset of Fig.~\ref{fig3} where the maximum of the dynamic susceptibility
is compared with the mean cluster size), and decays
after a time roughly of the order $\tau_b$. Only for times 
much smaller than $\tau_b$ the system behaves as if the bonds were
permanent. As the volume fraction increases, $\tau_b$ and
$\tau_\alpha$ become of the same order of magnitude, the plateau
disappears and $\chi_4(k_{min},t)$ exhibits a sharp peak similar to
those of glassy systems. We have also checked that for temperature
$T=0.25$, $\chi_4(k_{min},t)$ exhibits a small peak and no plateau.

\begin{figure}
\begin{center}
\includegraphics[width=7cm]{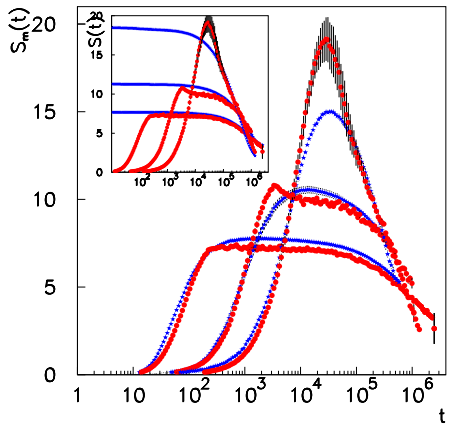}
\end{center}
\caption{(color online). {\bf Main frame}: The dynamic susceptibility,
$\chi_4(k_{min},t)$ (circles), compared with the time depending mean
cluster size of mobile particles, $S_m(t)$ (stars) for $T=~0.15$ and
$\phi=~0.05,~0.10,~0.12$. {\bf Inset}: $\chi_4(k_{min},t)$ (circles) compared 
with the time depending mean
cluster size, $S(t)$ (stars) for the same temperature and volume fractions.}
\label{fig4}
\vspace{-0.5cm}
\end{figure}

The long time behaviour of $\chi_4(k_{min},t)$ can be quantitatively
described in terms of a time dependent mean cluster size, $S(t)$.
The function $S(t)$ is the mean cluster size defined by persistent
bonds, that is bonds that are present at both time $0$ and time $t$
\cite{sciortino-water} (see inset of Fig. \ref{fig4}). 
To describe also the short time behaviour of 
$\chi_{4}(k_{min},t)$ we note that the main contribution to 
$\chi_{4}(k_{min},t)$ comes from the ``mobile'' particles \cite{nota_mobile}.
We therefore modify further the definition of the mean cluster size introducing 
a mean cluster size for ``mobile" particles, $S_m(t)$, where the
clusters are made as before by particles connected by persistent
bonds, but restricted to particles that in the time
interval $[0,t]$ have moved at least a distance $r_0$.
For each volume fraction, we fix $r_0$ so that the time at which $S_m(t)$ and
$\chi_{4}(k_{min},t)$ start to grow is the same.
We find that $r_0$ depends on the volume fraction roughly as $\phi^{-1/3}$.
In main frame of Fig.~\ref{fig4}, $S_m(t)$ is compared with $\chi_4(k_{min},t)$.
At least at very low volume fractions, the two quantities agree surprisingly 
well. At $\phi=0.10$ a deviation begins to appear, that
becomes manifest at $\phi=0.12$, in the shape of a peak that grows
to values higher than $S_m(t)$ or $S(t)$. This peak, which appears
when $\tau_\alpha$ and $\tau_b$ are of the same order of magnitude
shows that the contribution to dynamic correlations, comes not only
from the presence of long living clusters but also from the crowding
of the particles.

In conclusions we have shown that in a model for colloidal systems
at low volume fraction when the bond lifetime is much smaller than
the relaxation time, the dynamic heterogeneities can be described
in terms of time dependent clusters made of ``persistent" bonds.
We suggest that the analysis presented here should apply not only to
colloidal gelation at low volume fraction, but also to micellar
system, where a crossover from gel-like to glass-like behavior was
found experimentally and numerically \cite{mallamace et al}.
These findings offer a unique and coherent interpretation of dynamic
heterogeneities in gelling systems.
Finally, we also
suggest that the concept of time dependent cluster, considered
here, may be generalized to give a satisfactory definition of
dynamic heterogeneities also for Lennard-Jones or hard sphere
glasses.

The research is supported by the Marie Curie Reintegration
Grant MERG-CT-2004-012867 and EU Network Number MRTN-CT-2003-504712.


\begin{thebibliography}{99}
\bibitem{nota_bond}  Two particles
are considered linked if their relative distance is
smaller than the local maximum of the potential.
\bibitem{nota}
The non monotonic behaviour of $\tau_b$ is essentially due to the
presence of competitive interactions. After the formation of the
cluster phase, where all particles essentially feel only the
attractive interaction, the bond strength of further particles
decreases due to the effect of the long range repulsion.
\bibitem{nota_mobile} 
The distinction between mobile and quasi-frozen particles 
has been discussed in literature see for example \cite{kob,attr_sim,hurtado}.
\end{thebibliography}
\end{document}